\documentclass[ floatfix,twocolumn]{revtex4}
\usepackage{amsmath}
\usepackage{color}
\usepackage{amssymb}
\usepackage{graphicx}
\usepackage[utf8]{inputenc}
\usepackage[english, russian]{babel}
\usepackage{textcomp}
\usepackage{color}
\usepackage{esint}

\begin{document}

\title{ 
Toroid,  Altermagnetic, and Noncentrosymmetric ordering in metals}

\author{V.P.Mineev}
\affiliation{Landau Institute for Theoretical Physics, 142432 Chernogolovka, Russia}

\begin{abstract}
This article is dedicated to the 60-th anniversary of the Landau Institute for Theoretical Physics and presents a review of normal and superconducting properties of toroidal, altermagnetic, and noncentrosymmetric metals.
Metals with toroidal order are compounds not possessing symmetry in respect of space and time inversion but are symmetric in respect of the product of these operations.
An electric current propagating through samples of such a material causes its magnetisation.
Superconducting states in toroidal metals are a mixture of singlet and triplet  states. Superconductivity is gapless even in ideal crystals without impurities.
Altermagnets are antiferromagnetic metals 
that have a specific spin splitting of electron bands determined by time inversion in combinations with rotations and reflections of a crystal lattice. Similar splitting takes place  in metals whose symmetry does not have a spatial inversion operation. Both of these types of materials have an anomalous Hall effect. A current propagating through a noncentrosymmetric metal causes magnetization, but this is not the case in altermagnets. On the other hand, in altermagnets, there is a specific piezomagnetic Hall effect. Superconducting pairing in non-centrosymmetric metals occurs between electrons occupying states in one zone, whereas, in altermagnets, we are dealing with interband pairing, which is unfavorable for the formation of a superconducting state.

\bigskip


\end{abstract}

\date{\today}

\maketitle

{\bf CONTENTS}

\smallskip

$~~~~~~${\bf 1. Introduction}

$~~~~~~${\bf 2. Metals with toroid order}

$~~~~~~$A. Electron spectrum

$~~~~~~$B. Kramers degeneracy

$~~~~~~$C. Current in thermodynamic equilibrium

$~~~~~~$D. Zero-field current induced Hall effect

$~~~~~~$ {\bf 3. Superconducting states in torioid metals}

$~~~~~~$ A. Order parameter

$~~~~~~$ B. BCS theory

$~~~~~~~$ C. Free energy linear in order parameter gradients

$~~~~~~$ {\bf 4. Altermagnets and noncentrosymmetric metals}

$~~~~~~$ A. Electronic states

$~~~~~~$ B. Spin current in thermodynamic equilibrium

$~~~~~~$ C. Spin succeptibility

$~~~~~~$ D. Kinetic equation

$~~~~~~$ E. Conductivity

$~~~~~~$ G. Anomalous Hall effect 

$~~~~~~$ F. Piezomagnetic Hall effect 

$~~~~~~$ {\bf 5. Superconducting states in altermagnets  }

$~~~~~~$ {\bf References}

\section{Introduction}

Piezomagnetism and magnetoelectric effect in dielectric antiferromagnetic materials are well-known phenomena closely related to magnetic symmetry \cite{LL1957}. New interest in these phenomena has arisen recently in connection with the discovery of the first examples of metallic compounds with the same magnetic symmetry, but possessing new, sometimes unexpected physical properties. And, as is typical for the modern commercial style of writing scientific papers, a new sonorous terminology has appeared, designed to emphasize the significance of the authors' achievements. Thus, magnetoelectric metals began to be called {\bf metals with a toroidal order}. In turn, piezomagnetic metals were called {\bf altermagnets}. Somewhat earlier, the first metallic compounds were discovered whose symmetry does not contain the space inversion operation. They were called {\bf non-centrosymmetric metals}. This article presents an overview of the normal and superconducting properties of these three types of materials.

\section{Metals with toroid order}

Substanses with crystal symmetry which does not contain the operation of time reversal R as well as space inversion I but invariant in respect of its product IR called magneto-electrics. Landau and Lifshitz \cite{LL1957} have shown that if a crystal with such symmetry is placed in a constant magnetic (or electric) field, an electric (or magnetic) moment proportional to the field is produced in the crystal.
I.E.Dzyaloshinskii \cite {Dzyal1959} gave the first example of magnetoelectric antiferromagnetic Cr$_2$O$_3$.
It has the point symmetry group
\begin{equation}
{\bf D}_{3d}({\bf D}_3)=(E, C_3,C_3^2,3u_2,3\sigma_dR,2S_6R,IR)
\label{d}
\end{equation} 
 containing the product of time and space inversion, but does not include these operations separately. The corresponding thermodynamic potential invariant in respect to  operations (\ref{d}) is
 \begin{equation}
 \Phi_{em}=-\alpha_\perp(E_xH_x+E_yH_y)-\alpha_\parallel E_zH_z.
 \end{equation}
So, this material in external electric field acquires magnetisation
\begin{equation}
M_y=-\frac{\partial \Phi_{em}}{\partial H_y}=\alpha_\perp E_y.
\end{equation}

The magnetoelectric effect  in the antiferromagnet Cr$_2$O$_3$ was discovered by D.N.Astrov \cite {Astrov1960}. Despite of absence of magnetic moment this material is also exibit magneto-electric Kerr effect that is rotation of polarisation of light reflected from the crystal in respect of incident light polarisation. This birefringence is of opposite sign for magnetic domains related to each other by time reversal and can be used for observation of antiferromagnetic domains. The corresponding symmetry considerations was developed in the elegant paper by W.F.Brown et al \cite{Brown1963}, although the microscopic theory of this phenomenon  \cite{Muthukumar1995} and complete phenomenological treatment  \cite{Graham1997} appeared already after the effect was discovered experimentally \cite{Krichevtsov1993}.

\subsection{Electron spectrum}

Cr$_2$O$_3$ is antiferromagnetic dialectric. A metal with the same symmetry as Cr$_2$O$_3$ also possesses magnetoelecric properties.  Electron spectrum of such a metal  invariant in respect of all operations of the group ${\bf D}_{3d}({\bf D}_3)$
\begin{eqnarray}
\varepsilon_{\bf k}=\varepsilon^e_{\bf k}+\varepsilon^o_{\bf k},~~~~~~\varepsilon^e_{\bf k}=f(k_x^2+k_y^2,k_z^2),\\\varepsilon^o_{\bf k}=\gamma
(3k_y^2k_x-k_x^3)~~~~~~~
\label{epsilon}
\end{eqnarray}
consists from two parts even and odd in respect to  its argument ${\bf k}$.  
This is a general property of a metal with a symmetry that does not include the operations of time inversion $R$ and space inversion $I$ separately, but is invariant with respect to its product $IR$. Such metals are called {\bf metals with toroidal order} or simply {\bf toroids}. There is vast literature devoted to substances with toroidal order, see for instanse \cite{Kopaev2009,Hayami2014}.
Normal properties and superconducting states in toroids were discussed in the article \cite{Mineev2024}.

\begin{figure}
\includegraphics
[height=.2\textheight]
{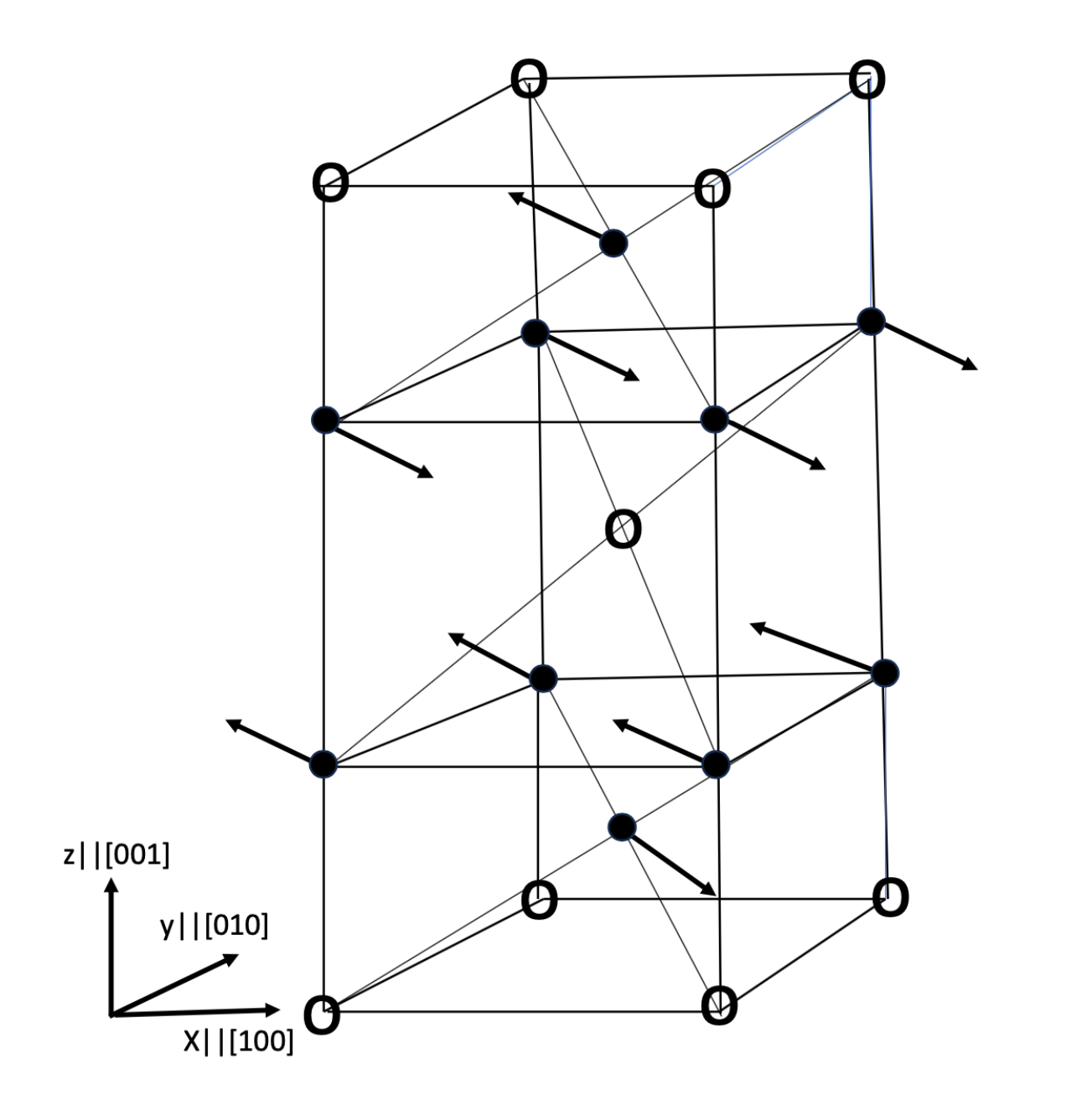}
 \caption{
Magnetic structure of Mn$_2$Au showing the order and orientation of the Mn ions magnetic moments (see the text). The small circles correspond to gold sites. }
\end{figure}

Recently there was discovered 
\cite{Fedchenko2022}  metallic compound  Mn$_2$Au with toroidal magnetic order.   Mn$_2$Au is collinear antiferromagnet with Neel vector parallel or antiparallel to [110] or $[1\bar10]$ directions. On the Fig1. is shown the magnetic structure of antiferromagnetic domain of this compound  
with the Neel vector parallel to $[1\bar10]$ direction.

Its symmetry group is
\begin{equation}
{\bf D}_{2h}({\bf C}_{2v})=(E,U_{xy},\sigma_h,\sigma_{x\bar y},RU_{x\bar y}, R\sigma_{xy},RC_{2z},RI).
\label{6}
\end{equation}
Here, the operations  $(E, U_{xy},\sigma_h,\sigma_{x\bar y})$ forming group ${\bf C}_{2v}$ are the operation of rotation on angle $\pi$ around axis   [110] and reflections in the planes passing through it and perpendicular to each other.
The electron spectrum invariant in respect of all operations of the group ${\bf D}_{2h}({\bf C}_{2v})$ is 
\begin{eqnarray}
\varepsilon_{\bf k}=\varepsilon^e_{\bf k}+\varepsilon^o_{\bf k},~~~~~~\varepsilon^e_{\bf k}=f(k_x^2+k_y^2,k_z^2),\\\varepsilon^o_{\bf k}=\gamma
(k_x+k_y).~~~~~~~
\label{epsilon}
\end{eqnarray}
The Fermi surface determined by equation
\begin{equation}
\varepsilon_{\bf k}=\varepsilon_F,
\end{equation}
is assymetrical because $\varepsilon_{\bf k} \ne \varepsilon_{-{\bf k}}$.

\subsection{Kramers degeneracy}

 The Hamiltonian in the Schr\"odinger equation for an electron in such a metal commutes with the product of time and space inversion operations $RI$.
 This means that to each energy $\varepsilon_{\bf k}$ corresponds two spinor eigen-functions $\psi_{\alpha}({\bf r})$
 and $RI \psi_{\alpha}({\bf r})$. They are orthogonal to each other. Indeed,
 the operation of the time reversal is
 $
 R_{\alpha\beta}=-i\sigma_{\alpha\beta}^yK_0
 $,
where $\sigma_{\alpha\beta}^y$ is the Pauli matrix, $K_0$ is the operation of complex conjugation, and 
\begin{eqnarray}
{\cal I}=\int d^3{\bf r}\left [\psi^\star_{\alpha}({\bf r})I R_{\alpha\beta}\psi_{\beta}({\bf r})\right]\nonumber\\=
\int d^3{\bf r}\left [\psi^\star_{\alpha}({\bf r})(-i)\sigma_{\alpha\beta}^y\psi^\star_{\beta}(-{\bf r})\right]=\nonumber\\
-\int d^3{\bf r}\left [\psi^\star_{\alpha}({\bf r})(-i)\sigma_{\beta\alpha}^y\psi^\star_{\beta}(-{\bf r})\right ]\nonumber\\=
-\int d^3{\bf r}\left [\psi^\star_{\beta}({\bf r})I R_{\beta\alpha}\psi_{\alpha}({\bf r})\right]=-{\cal I}.
\end{eqnarray}
Thus, ${\cal I}=0$. Hence, the Kramers degeneracy of each energy level takes place.

\subsection{Current in thermodynamic equilibrium}

Due to assymmetry of the energy spectrum toroid metals  possess nonzero electric current in thermodynamic equilibrium
\begin{equation}
{\bf j}=2e\int\frac{d^3k}{(2\pi)^3}\frac{\partial\varepsilon_{\bf k}}{\partial {\bf k}}f(\varepsilon_{\bf k}),
\label{curr}
\end{equation}
where $f(\varepsilon_{\bf k})=(\exp\frac{\varepsilon_{\bf k}-\mu}{T}+1)^{-1}$ is the Fermi distribution function. These currents remind
dissipationless diamagnetic currents flowing in metals in magnetic field.
Similar property takes place also in noncentrosymmetric metals supporting 
dissipationless spin currents in thermodynamic equilibrium.  This phenomenon we will discuss later.
 In real specimens with many antiferromagnetic domains
the currents  and corresponding magnetic moments space distribution acquire complex structure.

\subsection{Zero-field current induced Hall effect}

Toroid metals are magnetoelectrics. An electric field applied to such a metal induces  magnetization. For example, in the case of mono-domain
antiferromagnet Mn$_2$Au with structure shown in Fig. 1
the  thermodynamic potential invariant in respect to all operations enumerated in (\ref{6}) is
 \begin{equation}
 \Phi_{em}=-\alpha (E_{x\bar y}H_z+E_zH_{x\bar y}).
 \end{equation}
An electric field directed along $z$-axis causes magnetzation parallel or antiparallel to the direction of the Neel vector
\begin{equation}
M_{x\bar y}=\alpha E_z.
\end{equation}
One can  also say that an electric current $j_z=\rho^{-1}_z E_z$ along $z$-axis causes magnetisation
\begin{equation}
M_{x\bar y}=\alpha \rho_z j_z.
\end{equation}
As a result, an electric field arises in such a sample that is perpendicular to both the current and the induced magnetic moment
\begin{equation}
E_{xy}=\frac{1}{nec}\alpha \rho_z j^2_z.
\end{equation}
This is the current induced Hall effect in zero magnetic field. In multi-domain specimens the current induced magnetisation will have complex space distribution.

The effect of bulk magnetisation
induced by electric current has been observed in semiconducting tellurium \cite{Furukawa2017} and then in antiferromagnetic metallic compound UNi$_4$B \cite{Amitsuka2018,Izawa2022} where  the zero-field Hall effect was also registered \cite{Izawa2022}.

\section{Superconducting states in toroid metals}

\subsection{Order parameter}

Superconducting compounds with toroid symmetry are at the moment unknown. The theory of superconductivity for such type of substances will be presented here with hope on possible applications to be appear in future. In the absence of symmetry in respect of space inversion the superconducting order parameters  in toroid metals consist from sum of singlet and triplet parts
\begin{equation}
 \Delta_{{\bf k},\alpha\beta}=\Delta\Phi_{\alpha\beta}({\bf k})=\Delta \left [ \phi^s_{\bf k}i\sigma^y_{\alpha\beta}+(\mbox{\boldmath$\phi$}^t_{\bf k}\mbox{\boldmath$\sigma$}_{\alpha\gamma})i\sigma^y_{\gamma\beta}\right ].
\label{op}
 \end{equation}
Here, $\Delta$ is the coordinate dependent complex amplitude,
$ \mbox{\boldmath$\hat\sigma$}=(\hat\sigma^x,\hat\sigma^y,\hat\sigma^z)$ are the Pauli spin matrices. 
The functions $\phi^s_{\bf k}$ and 
$\mbox{\boldmath$\phi$}^t_{\bf k}$ correspond to representations of the symmetry group of concrete toroidal metal. For instance, in the case of single domain antiferromagnet with symmetry group (\ref{6}) the functions of irreducible
representations $\Gamma=A,B,C,D$ are presented in the table.

\begin{widetext}

 ~~~~~~~~~~~~~~~~~~~~~~~~~~~~~~~\begin{tabular}{|c|c|c|}
\hline
 $\Gamma$& $\phi^s_{\bf k}$&$\mbox{\boldmath$\phi$}^t_{\bf k}$ \\
 \hline
 ~A~& ~$ a_1(\hat k_x+\hat k_y)^2+a_2\hat k_z^2$~&~$ ia_3( \hat k_x-\hat k_y)\hat z$~\\
 \hline
 ~B~&~$b_1(\hat k_x+\hat k_y)\hat k_z$~&~$ ib_2(\hat k_x\hat y-\hat k_y\hat x)$~\\
 \hline
  ~C~&~$c_1(\hat k_x-\hat k_y)\hat k_z$~&~$ ic_2(\hat k_x+\hat k_y)(\hat x+\hat y)+ic_3(\hat k_x-\hat k_y)(\hat x-\hat y)+ic_4\hat k_z\hat z$~\\
  \hline
 ~D~&~$d_1(\hat k_x+\hat k_y)(\hat k_x-\hat k_y)$~&~$ id_2(\hat k_x+\hat k_y)\hat z$~\\
  \hline
 \end{tabular}
 
\end{widetext} 

Here,  $\hat k_x,\hat k_y,\hat k_z$ are the components of unit vector of momentum $\hat{\bf k}={\bf k}/|{\bf k}|$, and $\hat x,\hat y, \hat z$ are the unit vectors of directions in the spin space.

\subsection{BCS theory}

The BCS Hamiltonian has the standard form 
\begin{eqnarray}
H=H_0+H_{int}=\sum_{\bf k}(\xi_{\bf k}+\varepsilon^o_{\bf k})a^+_{{\bf k}\alpha}a_{{\bf k}\alpha}\nonumber\\
+\frac{1}{2}\sum_{{\bf k},{\bf k}^\prime}V_{\alpha\beta,\lambda\mu}({\bf k},{\bf k}^\prime)
a^+_{-{\bf k}\alpha}a^+_{{\bf k}\alpha}a_{{\bf k}^\prime\lambda}a_{-{\bf k}^\prime\mu}.
\label{H}
\end{eqnarray}
with only difference that kinetic energy has now even 
\begin{equation}
\xi_{\bf k}=\varepsilon^e_{\bf k}-\mu
\end{equation}
and odd $\varepsilon^o_{\bf k}$ in respect to momentum parts. 
In the pairing interaction 
\begin{equation}
V_{\alpha\beta,\lambda\mu}({\bf k},{\bf k}^\prime)=-V_\Gamma\Phi_{\alpha\beta}({\bf k})\Phi^\dagger_{\lambda\mu}({\bf k}^\prime)
\end{equation}
was left only term related to irreducible representation $\Gamma$ corresponding  to superconducting state with maximal critical temperature. 
After  usual mean field transformation the Hamiltonian acquires the following form
\begin{eqnarray}
H=\frac{1}{2}
\sum_{\bf k}(\xi_{\bf k}+\varepsilon^o_{\bf k})a^+_{{\bf k}\alpha}a_{{\bf k}\alpha}
-\frac{1}{2}\sum_{\bf k}(\xi_{-\bf k}+\varepsilon^o_{-\bf k})a_{-{\bf k}\alpha}a^+_{-{\bf k}\alpha}
\nonumber\\
+\frac{1}{2}\sum_{\bf k}\Delta_{{\bf k},\alpha\beta}a^+_{{\bf k}\alpha}a^+_{-{\bf k}\beta}+
\frac{1}{2}\sum_{\bf k}\Delta^\dagger_{{\bf k},\alpha\beta}a_{-{\bf k}\alpha}a_{{\bf k}\beta}~~~~\nonumber\\
+\frac{1}{2}\sum_{{\bf k}\alpha}(\xi_{-\bf k}+\varepsilon^o_{-\bf k})+\frac{1}{2}\sum_{\bf k}\Delta_{{\bf k},\alpha\beta}F^+_{{\bf k},\beta\alpha},~~~~~~
\label{mf}
\end{eqnarray}
 where the matrix of the order parameter
 \begin{equation}
 \Delta_{{\bf k},\alpha\beta}=-\sum_{{\bf k}^\prime}V_{\beta\alpha,\lambda\mu}({\bf k},{\bf k}^\prime) \langle a_{{\bf k}\lambda}a_{-{\bf k}\mu}   \rangle
  \label{Delta}
 \end{equation}
 is expressed through "anomalous average"
 \begin{equation}
 F_{{\bf k},\alpha\beta}=\langle a_{{\bf k}\alpha}a_{-{\bf k}\beta}   \rangle.  
 \end{equation}
 Here, $\langle ...\rangle$ means subsequent quantum mechanical and thermal averaging.

 More compact shape of  Eq.(\ref{mf}) is
 \begin{eqnarray}
 H=\frac{1}{2}\sum_{\bf k}\varepsilon_{{\bf k},ij}A^+_{{\bf k},i}A_{{\bf k},j}~~~~~\nonumber\\
 +\frac{1}{2}\sum_{{\bf k}\alpha}(\xi_{\bf k}-\varepsilon^o_{\bf k})
 +\frac{1}{2}\sum_{\bf k}\Delta_{{\bf k},\alpha\beta}F^+_{{\bf k},\beta\alpha}.
  \end{eqnarray}
 Here,
 the operators
 \begin{equation}
 A^+_{{\bf k},i}=(a^+_{{\bf k}\alpha,}a_{-{\bf k}\alpha}),~~~~~A_{{\bf k},i}=\begin{pmatrix} a_{{\bf k}\alpha}\\a^+_{-{\bf k}\alpha}   \end{pmatrix}
 \end{equation}
 and 
  \begin{equation}
 \varepsilon_{{\bf k},ij}=\begin{pmatrix}(\xi_{\bf k}+\varepsilon^o_{\bf k} )\delta_{\alpha\beta} &  \Delta_{{\bf k},\alpha\beta}\\
 \Delta^\dagger_{{\bf k},\alpha\beta}& (- \xi_{\bf k}+\varepsilon^o_{\bf k} ) \delta_{\alpha\beta} \end{pmatrix}.
\end{equation} 
 Diagonalising Hamiltonian by means the Bogolubov transformation
 \begin{eqnarray}
 A_{{\bf k},i}=U_{ij}B_{{\bf k},j},~~~B_{{\bf k},j}=\begin{pmatrix} b_{{\bf k}\alpha}\\b^+_{-{\bf k}\alpha}~~   \end{pmatrix},\nonumber\\
 U_{ij}=\begin{pmatrix}u_{{\bf k},\alpha\beta}&  v_{{\bf k},\alpha\beta}\\
 v^\dagger_{{\bf k},\alpha\beta}& - u_{{\bf k},\alpha\beta}\end{pmatrix},~~~~~~~~~~~
 \end{eqnarray}
 \begin{eqnarray}
 u_{{\bf k},\alpha\beta}=\frac{\xi_{\bf k}+E_{\bf k}^e}{\sqrt{(\xi_{\bf k}+E_{\bf k}^e)^2+\Delta^2_{\bf k}}}\delta_{\alpha\beta},\\
 v_{{\bf k},\alpha\beta}=\frac{\Delta_{\alpha\beta}({\bf k})}{\sqrt{(\xi_{\bf k}+E_{\bf k}^e)^2+\Delta^2_{\bf k}}},~~~~
 \end{eqnarray}
 \begin{equation}
 E_{\bf k}^e=\sqrt{\xi_{\bf k}^2+\Delta^2_{\bf k}},~~~~~~~~~\Delta^2_{\bf k}=\frac{1}{2}\Delta_{{\bf k},\alpha\beta}^\dagger\Delta_{{\bf k},\beta\alpha},
 \end{equation}
 we obtain
 \begin{equation}
 \frac{1}{2}\sum_{\bf k}\varepsilon_{{\bf k},ij}A^+_{{\bf k},i}A_{{\bf k},j}=\frac{1}{2}\sum_{\bf k}E_{{\bf k},ij}B^+_{{\bf k},i}B_{{\bf k},j},
 \end{equation}
 where
 \begin{equation}
 E_{{\bf k},ij}=\begin{pmatrix}(\varepsilon_{\bf k}^o+E_{\bf k}^e)\delta_{\alpha\beta} &  0\\
 0& (\varepsilon_{\bf k}^o-E_{\bf k}^e) \delta_{\alpha\beta} \end{pmatrix}.
 \end{equation}
 Thus, the energy of excitations is
 \begin{equation}
 E_{\bf k}=\varepsilon_{\bf k}^o+E_{\bf k}^e.
 \end{equation}
 The corresponding density of states is
 \begin{equation}
 N(E)=2\int\frac{d^3 {\bf k}}{(2\pi)^3}\delta(E-E_{\bf k}).
\end{equation} 
We see,  that near the surface  determined by equation $\xi_{{\bf k}}=0$ there is vast region where $E_{\bf k}< 0$, hence, a superconducting state is proved to be {\bf gapless } $$N(E=0)\ne 0.$$
This property of superconducting states in superconductors with toroidal order in particular means nonzero specific heat ratio $(C(T)/T)_{T\to 0}\ne 0$ in completely pure metal without impurities and crystal imperfections. 
 
The order parameter is determined by Eq.(\ref{Delta}).
 By application to this expression the Bogolubov transformation we obtain
 \begin{eqnarray}
 \Delta_{{\bf k},\alpha\beta}=-\int\frac{d^3 {\bf k}^\prime}{(2\pi)^3}V_{\beta\alpha,\lambda\mu}({\bf k},{\bf k}^\prime)
 \frac{1-f_{{\bf k}^\prime}-f_{{-\bf k}^\prime}}{2E_{{\bf k}^\prime}^e}\Delta_{{\bf k}^\prime,\lambda\mu}\nonumber\\
 =-\int\frac{d^3 {\bf k}^\prime}{(2\pi)^3}V_{\beta\alpha,\lambda\mu}({\bf k},{\bf k}^\prime)
 \frac{\tanh\frac{E_{{\bf k}^\prime}}{2T}+\tanh \frac{E_{{-\bf k}^\prime}}{2T}}
 {4E_{{\bf k}^\prime}^e}\Delta_{{\bf k}^\prime,\lambda\mu}.~
 \label{gap}
 \end{eqnarray}
 Here, we used the commutation rules of the operators $b_{{\bf k}\alpha}$, $b^+_{{\bf k}\alpha}$, the symmetry property
 \begin{equation}
  v_{{\bf k},\alpha\beta}=- v_{-{\bf k},\beta\alpha}
 \end{equation}
 and expressed the average $\langle b^+_{{\bf k}\alpha}b_{{\bf k}\beta}   \rangle=f_{\bf k}\delta_{\alpha\beta}$ through   the Fermi distribution function
 \begin{equation}
 f_{\bf k}=f(E_{\bf k})=\frac{1}{\exp((\varepsilon_{\bf k}^o+E_{\bf k}^e)/T)+1}.
 \end{equation}

 At $T\to T_c$ one can neglect $\Delta^2_{\bf k}$ in $E_{\bf k}^e$ in Eq.(\ref{gap}). Estimating the integral with logarithmic accuracy we come to
 the expression for critical temperature similar to usual BCS formula
 \begin{equation}
 T_c\approx \varepsilon_0\exp\left (-\frac{1}{\tilde N_0V_{\Gamma}}\right ),
 \end{equation}
where $\varepsilon_0$ is a cut-off  for energy of pairing interaction and $\tilde N_0$ is the normal density of states averaged over the Fermi surface with a weight corresponding to the angular dependent functions of given irreducible representation.

\subsection{Free energy linear in order parameter gradients}

Let us now discuss the possible peculiar property of inhomogenious state in  superconductors with toroidal symmetry.The expression for the superconducting current
\begin{equation}
{\bf j}=-\frac{2e}{\hbar}K\left[\Delta^\star(-i\nabla+\frac{2e}{\hbar c}{\bf A})\Delta+c.c.\right]
\end{equation}
changes its sign under the time reversal  R as well under the space inversion I, but it is invariant in respect to the product of this operations IR.
Thus, the current has the toroid symmetry. Hence, one can expect, as this was claimed for instance in \cite{Kopaev2009}, the existence of the linear in gradients term 
\begin{equation}
F_\nabla=C_ij_i
\label{F}
\end{equation}
in superconducting free energy density specific for the metals with toroid symmetry. The direction of vector ${\bf C}$ is determined by the direction of Neel vector of toroid antiferromagnet.

To verify   this property let us consider
the superconducting free energy quadratic in respect of the order parameter 
\begin{eqnarray}
{\cal F}=\frac{1}{2V}\int\frac{d^3{\bf q}}{(2\pi)^3}\Delta^\star({\bf q} )\Delta({\bf q})
 -\frac{T}{2}\sum_{\omega}\int\frac{d^3{\bf q}}{(2\pi)^3}\int\frac{d^3{\bf k}}{(2\pi)^3}
\nonumber\\
\times \Delta^\star_{{\bf k},\alpha\beta}({\bf q})G(-{\bf k}+{\bf q}/2,-\omega_n)
G({\bf k}+{\bf q}/2,\omega_n)\Delta_{{\bf k},\beta\alpha}({\bf q}),~~
\end{eqnarray}
where
\begin{equation}
G({\bf k},\omega_n)=\frac{1}{i\omega-\xi_{\bf k}-\varepsilon^o_{\bf k}}
\end{equation}
is the normal state electron Green function, $\omega_n=\pi T(n+1/2)$ is the Matsubara frequency. Omitting simple but cumbersome calculations, we only indicate that after performing the  summation over frequencies  followed by  the decomposing  of the sub-integral expression in powers of 
$\frac{\partial\xi_{\bf k}}{\partial{\bf k}}{\bf q}$ and $\frac{\partial\varepsilon^o_{\bf k}}{\partial{\bf k}}{\bf q}$ 
the integral over angles of momentum ${\bf k}$
of the  linear in ${\bf q}$ part turns out to be equal to zero. 
This means that  the term (\ref{F}) vanishes identically. 

\section{Altermagnetic and Noncentrosymmetric  metals}

There is another type of magnetic structures in which  the magnetic symmetry group does not contain the time reversal R by itself but this operation enters only in combination with rotations or reflections, or else is not present at all. Consequently such substances, in general, are capable of posessing piezomagnetic properties \cite{Tavger1956,LL1957,Dzyal1957}. Piezomagnetism was discovered in antiferromagnetic fluorides of cobalt CoF$_2$ and manganese MnF$_2$ by A.S.Borovik-Romanov \cite{Borovik1960}.
These substaces have a simple tetragonal lattice and the symmetry of space group ${\bf D}_{4h}^{14}$. In their unit cell there are two metallic ions in positions (000) and $(\frac{1}{2},\frac{1}{2},\frac{1}{2})$. The magnetic structure has been determined neutronographically by R.A.Erickson \cite{Ericson1953}. (see Fig2.)

The group of symmetry of CoF$_2$ and MnF$_2$ is
\begin{eqnarray}
{\bf D}_{4h}({\bf D}_{2h})=(E,C_{2},2U_2t,\sigma_h, 2\sigma_vt, I,\nonumber\\2C_{4z}Rt,2U'_2R,2\sigma'_vR,2C_{4}\sigma_hRt).
\label{sym1}
\end{eqnarray}
Here we use the same notations for the operations of rotations and reflections as in the textbook \cite{QM}. For example, $2U'_2R$ - rotations on angle $\pi$ around [110] or $[1\bar 10]$ axis accompanied by operation of time reversal $R$. 
The crystal symmetry of these substances is nonsymmorphic and some of operations enumerated in (\ref{sym1})  are accompanied by the  shift on half period $t=t_{1/2}=(a,a,c)/2$ along the prism diagonal.

\begin{figure}
\includegraphics
[height=.2\textheight]
{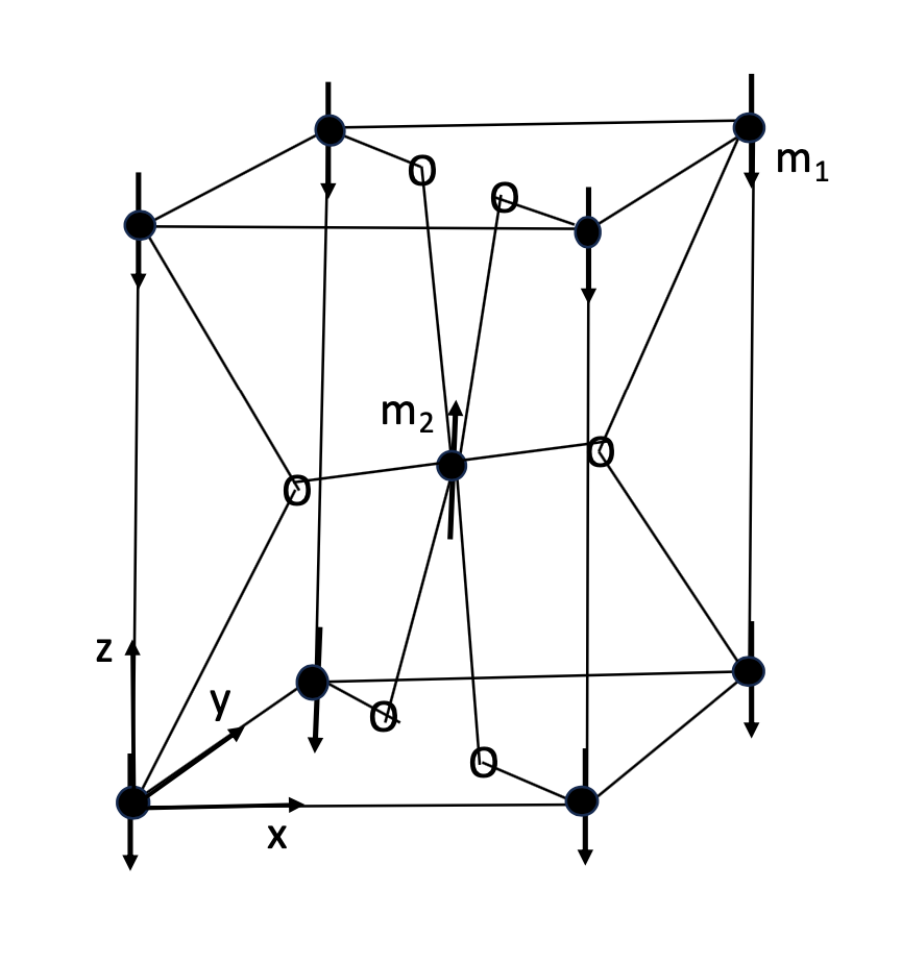}
 \caption{Magnetic structure of MnF$_2$ showing the order and orientation of the Mn ions magnetic moments. The small circles correspond to fluorine sites. }
\end{figure}

On the large scale in comparison with interatomic distances the operation $t$-shift plays no role and the essential symmetry is only in respect to rotations and reflections in combination with time reversal R.
The piezomagnetic thermodynamic potential invariant in respect of all these operations is
\begin{equation}
\Phi_{pm}=-\lambda_1(\sigma_{xz}H_y+\sigma_{yz}H_x)-\lambda_2\sigma_{xy}H_z
\end{equation}
and corresponding additional magnetisation arising under application of shear stress  $\sigma_{xz}$  is
\begin{equation}
M_y=-\frac{\partial \Phi_{pm}}{\partial H_y}=\lambda_1\sigma_{xz}.
\end{equation}
This effect was measured and reported in \cite{Borovik1960}.

Both CoF$_2$ and MnF$_2$ are dielectric antiferromagnets. 
The same crystallographic structure and antiferromagnetic order has metallic compound RuO$_2$ determined by Z.H.Zhu et al \cite{Zhu2019} by means resonant X-ray scattering.
The energy of electron as a function of momentum in a metal with structure symmetric in respect of all the operations pointed in Eq.(\ref{sym1}) has the 
following form
\begin{equation}
\varepsilon_{\alpha\beta}=\varepsilon_{\bf k}\delta_{\alpha\beta}+\mbox{\boldmath$\gamma$}_{\bf k}\mbox{\boldmath$\sigma$}_{\alpha\beta},
\label{alt}
\end{equation}
\begin{eqnarray}
\mbox{\boldmath$\gamma$}_{\bf k}=\gamma_1 \sin(k_zb) \left[\sin(k_ya)\hat x+\sin(k_xa)\hat y\right]\nonumber\\
+\gamma_2\sin (k_xa)\sin(k_ya)\hat z,
\label{alt1}
\end{eqnarray}
where $\varepsilon=\varepsilon({\bf k})$ is
translation invariant even function with symmetry  ${\bf D}_{4h}({\bf D}_{2h})$
 and
$\mbox{\boldmath$\sigma$}=(\sigma_x,\sigma_y,\sigma_z)$ are the Pauli matrices. Here, we have  taken into account that to the  operation $t_{1/2}$ in coordinate space corresponds
shift $\pi(1/a,1/a,1/b)$ on half basis vector in the reciprocal space. 
The equation (\ref{alt1}) defining the vector  $\mbox{\boldmath$\gamma$}_{\bf k}$ is the simplest possible expression that has the necessary symmetry properties.

In general, the electron spectrum of a metal such that its group of symmetry $G$ ( magnetic class) contains the operation of time reversal only in 
combination with rotations or reflections has the form Eq.(\ref{alt})
invariant in respect of all operations of the group  $G$. There is subclass of  these type metals such that the angular average 
\begin{equation}
\int\frac{d\Omega_{\bf k}}{4\pi}\mbox{\boldmath$\gamma$}_{\bf k}=0.
\label{am1}
\end{equation}
These type of metals looking like antiferromagnets in reciprocal space are called {\bf altermagnets}. 
The electron spin ordering in altermagnets determined by exchange interacton is  in general  non-collinear.

In some cases one must take into account interband spin-orbit interaction and work with  more   general $4\times4$ matrix electron spectrum
\begin{eqnarray}
\hat{\cal E}_{\bf k}=(\varepsilon_{1{\bf k}}+\mbox{\boldmath$\gamma$}_{1{\bf k}}\mbox{\boldmath$\sigma$})(\tau_0+\tau_3)/2\nonumber\\+
i\tau_2\mbox{\boldmath$\varphi$}_{\bf k}\mbox{\boldmath$\sigma$}+
(\varepsilon_{2{\bf k}}+\mbox{\boldmath$\gamma$}_{2{\bf k}}\mbox{\boldmath$\sigma$})(\tau_0-\tau_3)/2.
\label{four}
\end{eqnarray}
Here $\tau_0,\tau_1, \tau_2,\tau_3$ are the band Pauli matrices. This form of spectrum is important in study of anomalous Hall effect in altermagnets (see below).

\subsection{Electronic states}

In the subsequent text  we will work with simple $2\times 2$ matrix spectrum
(\ref{alt}) which has the same form as in noncentrosymmetric metals 
\begin{equation}
\hat\varepsilon({\bf k})=\varepsilon_{\bf k}\sigma_0+\mbox{\boldmath$\gamma$}_{\bf k}\mbox{\boldmath$\sigma$}.
\label{11}
\end{equation} 
See for example \cite{Samokhin2008} and references therein.
Thus, all the calculations for these different types of metals look identical, but one must remember that in altermagnets vector 
$
\mbox{\boldmath$\gamma$}_{-{\bf k}}=\mbox{\boldmath$\gamma$}_{\bf k}
$
is even function of ${\bf k}$, whereas in noncentrosymmetric metals it is odd one $
\mbox{\boldmath$\gamma$}_{-{\bf k}}=-\mbox{\boldmath$\gamma$}_{\bf k}
$. Scalar part of spectrum $\varepsilon_{\bf k}=\varepsilon_{-{\bf k}}$ is even in both cases.

The eigenvalues of the matrix (\ref{11}) are
\begin{equation}
    \varepsilon_{+}({\bf k})=\varepsilon+\gamma,~~~~~~~~ \varepsilon_{-}({\bf k})=\varepsilon-\gamma,
\label{e3}
\end{equation}
where $\gamma=|\mbox{\boldmath$\gamma$}_{\bf k}|$.
The corresponding eigenfunctions are given by
\begin{eqnarray}
\Psi^+_\alpha({\bf k})=\frac{1}{\sqrt{2\gamma(\gamma+\gamma_z)}}\left (\begin{array} {c}
\gamma+\gamma_z\\
\gamma_+
\end{array}\right),\nonumber\\
~~~~~~~~~~~~\Psi^-_\alpha({\bf k})=\frac{t_+^\star}{\sqrt{2\gamma(\gamma+\gamma_z)}}
\left(\begin{array} {c}
-\gamma_-\\
\gamma+\gamma_z
\end{array}\right),
\label{ps}
\end{eqnarray}
where $\gamma_\pm=\gamma_x\pm i\gamma_y$ and $t_+^\star=-\frac{\gamma_+}{\sqrt{\gamma_+\gamma_-}}$.
The eigenfunctions obey the orthogonality conditions
\begin{equation}
\Psi^{\lambda_1\star}_\alpha({\bf k})\Psi^{\lambda_2}_\alpha({\bf k})=\delta_{\lambda_1\lambda_2},~~~~~~~
\Psi^\lambda_{\alpha}({\bf k})\Psi^{\lambda\star}_{\beta}({\bf k})=\delta_{\alpha\beta}.
\label{ort}
\end{equation}
Here, a summation over the repeating  spin $\alpha=\uparrow,\downarrow$
or band $\lambda=+,-$ indices is implied. 

In altermagnets as in noncentrosymmetric metals the eigen functions from different bands  transform to each other by operation 
of time inversion 
$$-i(\sigma_y)_{\alpha\beta}K_0\Psi_{\beta}^+({\bf k})\propto\Psi_{\alpha}^-({\bf k}),$$
where $K_0$ is the operation of complex conjugation.
  Thus, the Kramers degeneracy is lifted.

There are two Fermi surfaces with different Fermi momenta ${\bf k}_{F\pm}$.  They are determined by the equations
\begin{equation}
\label{e4}
    \varepsilon_{\pm}({\bf k})=\mu,
\end{equation}
and the Fermi velocities are given by the derivatives
\begin{equation}
{\bf v}_{F\pm}=\frac{\partial(\varepsilon_{\pm}({\bf k})}{\partial {\bf k}}|_{k=k_{F\pm}}.
\end{equation}

\subsection{Spin current in thermodynamic equilibrium}

The  density of spin current  is
 \begin{equation}
 {\bf j}_{i}=\int \frac{d^3k}{(2\pi)^3}\mbox{\boldmath$\sigma$}_{\alpha\beta}\frac{\partial \varepsilon_{\beta\gamma}({\bf k})}{\partial k_i}n_{\gamma\alpha}({\bf k},\omega).
 \label{sc}
\end{equation}
The matrix of the equilibrium electron distribution function is
\begin{equation}
n_{\alpha\beta}
=\frac{n_++n_-}{2}\delta_{\alpha\beta}+\frac{n_+-n_-}{2\gamma} 
\mbox{\boldmath$\gamma$}\cdot \mbox{\boldmath$\sigma$}_{\alpha\beta},
\label{eqv}
\end{equation}
where $n_\lambda=\left(e^{\frac{\varepsilon_\lambda-\mu}{T}}+1\right)^{-1}$ is the Fermi distribution function.

The integral (\ref{sc})  in altermagnets is equal to zero. However, in noncentrosymmetric metals there is nonzero spin current density in thermodynamic equilibrium \cite{Rashba2003,Rashba2004}
\begin{equation}
 {\bf j}_{i}=\int \frac{d^3k}{(2\pi)^3}\left [ \frac{\partial \mbox{\boldmath$\gamma$}}{\partial k_i}(n_++n_-)+\frac{\partial\varepsilon_{\bf k}}{\partial k_i}
(n_+-n_-)
\frac{\mbox{\boldmath$\gamma$}}{\gamma}\right]
\end{equation}
The presence of dissipationless spin currents is the property of noncentrosymmetric metals similar to  the presence of electric currents in thermodynamic equilibrium in a metal with toroid order given by Eq.(\ref{curr}).

\subsection{Spin susceptibility}

The spin quantisation axis is  given by the unit vector $\hat {\mbox{\boldmath$\gamma$}}=\mbox{\boldmath$\gamma$}/|\mbox{\boldmath$\gamma$}|$.
The  projections of the electron spins in two bands on the $\hat {\mbox{\boldmath$\gamma$}}$ direction have opposite signs 
\begin{equation}
(\hat {\mbox{\boldmath$\gamma$}}_{\bf k}\mbox{\boldmath$\sigma$}_{\alpha\beta})\Psi^{\pm}_{\beta}({\bf k})=\pm\Psi^{\pm}_{\alpha}({\bf k}).
\end{equation}

In an external magnetic field the matrix of the electron  energy is
\begin{equation}
\hat\varepsilon({\bf k})=\varepsilon_{\bf k}\sigma_0+\mbox{\boldmath$\gamma$}_{\bf k}\mbox{\boldmath$\sigma$}-{\bf h}\mbox{\boldmath$\sigma$}.
\label{hmatrix}
\end{equation}
The field is here written as ${\bf h}=\mu_B{\bf H}$.
The band energies are now given by
\begin{equation}
 \varepsilon_{\lambda,{\bf h}}({\bf k})=\varepsilon_{\bf k}+\lambda
|\mbox{\boldmath$\gamma$}_{\bf k}-{\bf h}|, ~~~\lambda=\pm.
\end{equation}
Along with the changes of the band energies, the spin quantisation axis also deviates from its zero field direction
\begin{equation}
\hat {\mbox{\boldmath$\gamma$}}_{\bf k}~~~\to~~~\hat {\mbox{\boldmath$\gamma$}}_{\bf h}({\bf k})=
\frac{\mbox{\boldmath$\gamma$}_{\bf k}-{\bf h}}{|\mbox{\boldmath$\gamma$}_{\bf k}-{\bf h}|}.
\end{equation}
The magnetic moment is written as
\begin{equation}
{\bf M}=\mu_B\int \frac{d^3k}{(2\pi)^3}\hat {\mbox{\boldmath$\gamma$}}_{\bf h}({\bf k})\left[n( \varepsilon_{+,{\bf h}}({\bf k}))
-n( \varepsilon_{-,{\bf h}}({\bf k}))\right],
\end{equation}
where $n(\varepsilon_\lambda)=\left(e^{\frac{\varepsilon_\lambda-\mu}{T}}+1\right)^{-1}$ is the Fermi distribution function.
Taking the term of the first order in magnetic field
we obtain for the magnetic susceptibility
\begin{eqnarray}
\chi_{ij}=-\mu_B^2\int \frac{d^3k}{(2\pi)^3}\left\{\hat {\mbox{\boldmath$\gamma$}}_i\hat {\mbox{\boldmath$\gamma$}}_j
\left[\frac{\partial n( \varepsilon_{+})}{\partial\varepsilon_+}
+\frac{\partial n( \varepsilon_{-})}{\partial\varepsilon_-}\right]+\right.\nonumber\\
\left.+(\delta_{ij}-\hat {\mbox{\boldmath$\gamma$}}_i\hat {\mbox{\boldmath$\gamma$}}_j )\frac{n( \varepsilon_{+})
-n( \varepsilon_{-})}{|\mbox{\boldmath$\gamma$}|}\right\}.~~~~~~~~
\label{chi}
\end{eqnarray}
The first term under the sign of integration contains the derivatives of the jumps in the Fermi distributions ${\partial n( \varepsilon_{\pm})}/{\partial\varepsilon_{\pm}}=-\delta(\varepsilon_\pm-\mu)$. The second one originates from the deviation in the spin quantisation direction for the quasiparticles filling the states between the Fermi surfaces of two bands.
Thus, magnetic moment arising  in altermagnets in external magnetic field is determined by the same formula as in noncentrosymmetric metals  \cite{Mineev2010}.

\subsection{Kinetic equation}

In the band representation the equilibrium distribution function (\ref{eqv})  is given by the diagonal matrix
\begin{equation}
n_{\lambda_1\lambda_2}=\Psi^{\lambda_1\star}_{\alpha}({\bf k})n_{\alpha\beta}\Psi^{\lambda_2}_{\beta}({\bf k})=\left (\begin{array} {cc} n({\varepsilon}_+)&0\\0&n({\varepsilon}_-)  \end{array}\right)_{\lambda_1\lambda_2}.
\end{equation}

The Hermitian matrices of the non-equilibrium distribution functions in the band and spin representations  related by
\begin{equation}
f_{\lambda_1\lambda_2}({\bf k})=\Psi^{\lambda_1\star}_{\alpha}({\bf k})f_{\alpha\beta}\Psi^{\lambda_2}_{\beta}({\bf k}).
\label{f}
\end{equation}

The kinetic equation for the electron distribution function in noncentrosymmetric metals has been obtained in \cite{Mineev2019}  from  the general matrix quasi-classic kinetic equation derived by V.P.Silin \cite{Silin1957}. 
In presence of  electric field ${\bf E}$ the 
the linearised matrix  kinetic equation  for the Fourier amplitudes of deviation of distribution function from equilibrium  $$g_{\lambda_1\lambda_2}({\bf k},\omega)=f_{\lambda_1\lambda_2}({\bf k})-n_{\lambda_1\lambda_2}
$$ 
is
\begin{eqnarray}
 e\left (\begin{array} {cc}({\bf v}_{+}{\bf E}) \frac{\partial n_+}{\partial \varepsilon_+}&({\bf w}_{\pm}{\bf E})(n_--n_+)
\\
({\bf w}_{\mp}{\bf E})(n_+-n_-))&({\bf v}_{-}{\bf E})\frac{\partial n_-}{\partial \varepsilon_-}
 \end{array}\right)~~~~~~~\nonumber\\+
 \left(
\begin{array} {cc}0&i(\varepsilon_--\varepsilon_+)g_{\pm}({\bf k})\\
i(\varepsilon_+-\varepsilon_-)g_{\mp}({\bf k})&0
 \end{array}\right)=\hat I.~~
 \label{eqv1}
\end{eqnarray}
Here, we put for brevity $n(\varepsilon_+)=n_+, ~n(\varepsilon_-)=n_-$. The quantities
\begin{eqnarray}
{\bf w}_{\pm}({\bf k})=
\Psi^{+\star}_{\alpha}({\bf k})\frac{\partial \Psi^{-}_{\alpha}({\bf k})}{\partial{\bf k}}~~~~~~~~\nonumber\\=\frac{t_+^\star}{2\gamma}\left (-\frac{\partial \gamma_-}{\partial{\bf k}}+\frac{\gamma_-}{\gamma+\gamma_z}\frac{\partial(\gamma+\gamma_z)}{\partial{\bf k}}\right),~~~~~
\label{vel}
\end{eqnarray}
are {\bf the interband} Berry connections,
$$
{\bf w}_{\mp}=-{\bf w}_{\pm}^\star.
$$
 Unlike to the group velocities ${\bf v}_+$,  ${\bf v}_-$, the dimensionality of the Berry connections ${\bf w}_\pm$
and ${\bf w}_\mp$ is $1/k$. 

$\hat I$ is the matrix integral of scattering. In Born approximation the collision integral $I_{\lambda_1\lambda_2}$ for electron scattering on impurities (see Appendix A in the paper \cite{Mineev2019}) is 
\begin{eqnarray}
I^i_{\lambda_1\lambda_2}({\bf k})=2\pi n_{imp}\int\frac{d^3k^\prime}{(2\pi)^3}|V({\bf k}-{\bf k}^\prime)|^2~~~~~~~~~~~~~~~~\nonumber\\
\times\left\{O_{\lambda_1\nu}({\bf k},{\bf k}^\prime)
\left [ g_{ \nu\mu}({\bf k}^\prime)O_{\mu\lambda_2}({\bf k}^\prime,{\bf k})\right.\right.~~~~~~~~~~~~~~~\nonumber\\
\left.\left.-O_{\nu\mu}({\bf k}^\prime,{\bf k})
  g_{ \mu\lambda_2}({\bf k}) \right ]\delta(\varepsilon^\prime_\nu
  -\varepsilon_{\lambda_2})\right.~~~~~~~~~~~~~~~~~~\nonumber\\ +
 \left. \left[O_{\lambda_1\nu}({\bf k},{\bf k}^\prime)g_{ \nu\mu}({\bf k}^\prime)\right.\right.~~~~~~~~~~~~~~~~~~~~~~~~~~~~~~~\nonumber\\-\left.\left.g_{ \lambda_1\nu}({\bf k})O_{\nu\mu}({\bf k},{\bf k}^\prime)
  \right ]O_{\mu\lambda_2}({\bf k}^\prime,{\bf k})\delta(\varepsilon^\prime_\mu-\varepsilon_{\lambda_1})\right \}.~
  \label{matrix1}
\end{eqnarray}
Here, we introduced notations $\varepsilon_{\lambda_1}=\varepsilon_{\lambda_1}({\bf k}),~ \varepsilon_{\mu}^\prime=\varepsilon_{\mu}({\bf k}^\prime)$ etc,
\begin{equation}
O_{\lambda_1\lambda_2}({\bf k},{\bf k}^\prime)=\Psi^{\lambda_1\star}_\sigma({\bf k})\Psi^{\lambda_2}_\sigma({\bf k}^\prime)
\end{equation}
such that
$$
O_{\lambda_1\lambda_2}({\bf k},{\bf k}^\prime)=O^\star_{\lambda_2\lambda_1}({\bf k}^\prime,{\bf k}).
$$
The expression for collision integral for electro-electron scattering one can find in the Appendix B in the paper \cite{Mineev2019}.

\subsection{Conductivity}

If the energy of band splitting exceeds the electron-
impurity scattering rate
\begin{equation}
v_F(k_{F-}-k_{F+})\gg1/\tau_{i}
\end{equation}
one can neglect by the collision integrals 
in the off-diagonal terms of matrix kinetic equation  (\ref{eqv1}) and use
the collision-less
solution  for the off-diagonal terms of the matrix distribution function 
\begin{eqnarray}
g_{\pm}=\frac{e({\bf w}_{\pm}{\bf E})(n_--n_+)}{-i(\varepsilon_--\varepsilon_+)},
\label{C}\\
g_{\mp}=\frac{e({\bf w}_{\mp}{\bf E})(n_+-n_-)}{-i(\varepsilon_+-\varepsilon_-)}.
\label{D}
\end{eqnarray}
There was shown that  in stationary case this type of the off-diagonal terms do not produce  a contribution to the electric current \cite{Mineev2019}.
On the other hand, substitution of these expressions to the diagonal parts of collision-integral matrices  (\ref{matrix1})
allows to neglect in them by all the terms containing off-diagonal elements of distribution function. These terms are $v_F(k_{F-}-k_{F+})\tau_i>>1$ times smaller than the terms
with diagonal elements.
Then the system Eq.({\ref{eqv1}) for
\begin{equation}
g_{\alpha\beta}({\bf k})=
\left (\begin{array} {cc} g_+({\bf k})&0\\0&g_-({\bf k})  \end{array}\right)_{\alpha\beta}
\end{equation}
acquires the following form:
\begin{equation}
({\bf v}_{+}{\bf E}) \frac{\partial n({\varepsilon}_+)}{\partial \varepsilon_+}=I_+^i,
\label{k}
\end{equation}
\begin{equation}
({\bf v}_{-}{\bf E}) \frac{\partial n({\varepsilon}_-)}{\partial \varepsilon_-}=I_-^i,
\label{l}
\end{equation}
where
\begin{eqnarray}
I_+^i=4\pi n_i\int\frac{d^3k}{2\pi^3}|V({\bf k}-{\bf k}^\prime)|^2\times\nonumber\\
\times\left\{
O_{++}({\bf k}{\bf k}^\prime)O_{++}({\bf k}^\prime{\bf k})[g_+({\bf k}^\prime)-g_+({\bf k})]\delta(\varepsilon_+^\prime-\varepsilon_+)\right.\nonumber\\
\left.+
O_{+-}({\bf k}{\bf k}^\prime)O_{-+}({\bf k}^\prime{\bf k})[g_-({\bf k}^\prime)-g_+({\bf k}))]\delta(\varepsilon_-^\prime-\varepsilon_+)
\right \},~~
\label{34}\\
I_-^i=4\pi n_i\int\frac{d^3k}{2\pi^3}|V({\bf k}-{\bf k}^\prime)|^2\times\nonumber\\
 \times\left\{
O_{--}({\bf k}{\bf k}^\prime)O_{--}({\bf k}^\prime{\bf k})[g_-({\bf k}^\prime)-g_-({\bf k})]\delta(\varepsilon_-^\prime-\varepsilon_-)\right.\nonumber\\+
\left.O_{-+}({\bf k}{\bf k}^\prime)O_{+-}({\bf k}^\prime{\bf k})[g_+({\bf k}^\prime)-g_-({\bf k}))]\delta(\varepsilon_+^\prime-\varepsilon_-)
\right \}.~~
\label{35}
\end{eqnarray}
Thus, we came to the system of two equations coupled through the collision integrals containing intraband and as well interband electron scattering terms.
One can  search 
  the solution of Eqs. (\ref{34}), (\ref{35})
  in the following form 
 \begin{eqnarray}
g_+=-e
\tau_+\frac{\partial n_+}{\partial \xi_+}({\bf v}_+{\bf E}),~~~
g_-=-e\tau_-\frac{\partial n_-}{\partial \xi_-}({\bf v}_-{\bf E})
,
\label{sol}
\end{eqnarray}
where the scattering times $\tau_+$, $\tau_-$ are even  functions of wave vector. They  should be found as solution of equations (\ref{34}),(\ref{35}).

  The electric  current density is
 \begin{equation}
 {\bf j}=e\int \frac{d^3k}{(2\pi)^3}\frac{\partial \varepsilon_{\alpha\beta}({\bf k})}{\partial {\bf k}}g_{\beta\alpha}({\bf k},\omega).
 \end{equation}
Transforming it to the band representation we obtain
\begin{eqnarray}
 {\bf j}=e\int \frac{d^3k}{(2\pi)^3} \Psi_\alpha^{\lambda_1\star}({\bf k})\frac{\partial \varepsilon_{\alpha\beta}({\bf k})}{\partial {\bf k}}
 \Psi_{\beta}^{\lambda_2}({\bf k})\nonumber\\
 \times\Psi_{\gamma}^{\lambda_2\star}({\bf k})
 g_{\gamma\delta}({\bf k},\omega)\Psi_{\delta}^{\lambda_1}({\bf k})\nonumber\\
   =e\int \frac{d^3k}{(2\pi)^3} \left \{
 \frac{\partial \varepsilon_{\lambda_1\lambda_2}({\bf k})}{\partial {\bf k}}+\left [ {\bf w}_{\lambda_1\lambda_3},  \varepsilon_{\lambda_3\lambda_2} \right]\right\}
 g_{\lambda_2\lambda_1}({\bf k})
,~~
\label{current}
\end{eqnarray}
where $\left [\dots,\dots\right ]$ is the commutator.
 Performing matrix multiplication we obtain
\begin{eqnarray}
{\bf j}=e\int \frac{d^3k}{(2\pi)^3} \left [ {\bf v}_+g_++{\bf v}_-g_-\right.\nonumber\\+\left.+({\bf w}_{\pm}g_{\mp}- {\bf w}_{\mp}g_{\pm})(\varepsilon_--\varepsilon_+)  \right ].
\label{cur}
\end{eqnarray}
In neglect off-diagonal terms of distribution function and substituting solutions Eq.(\ref{sol})
we obtain the expression
\begin{equation}
{\bf j}=-e^2\int \frac{d^3k}{(2\pi)^3} \left [ \tau_+\frac{\partial n_+}{\partial \xi_+}{\bf v}_+({\bf v}_+{\bf E})+
\tau_-\frac{\partial n_-}{\partial \xi_-}{\bf v}_-({\bf v}_-{\bf E})
\right ].
\label{cur1}
\end{equation}
determining the conductivity due to electron scattering on impurities. The corresponding derivation of conductivity determined by joint processes
of scattering on impurities and electron-electron scattering is derived in 
the paper \cite{Mineev2021}.

\subsection{Magnetoelectric effect}

Altermagnets are invariant in respect of space inversion, hence the external electric field does not cause magnetisation to appear in them. On the contrary noncentrosymmetric metals placed in an electric field possess  magnetolectric effect. In semiconductors this effect was predicted long ago by E.L. Ivchenko and G.E. Pikus \cite{Ivchenko1978} and reviewed in the recently published paper \cite{Ganichev2024}.  The magnetoelecricity in 2D metal  with the Rashba spin-orbit interaction    was considered first by V.M.Edelstein \cite{Edelstein1990}. More general treatment has been developed recently in the paper \cite{MineevLT2024}.

The density of magnetisation 
\begin{equation}
{\bf M}=\int\frac{d^3{\bf k}}{(2\pi)^3}\mbox{\boldmath$\sigma$}_{\alpha\beta}g_{\beta\alpha}=\int\frac{d^3{\bf k}}{(2\pi)^3}\mbox{\boldmath$\sigma$}_{\lambda_1\lambda_2}g_{\lambda_2\lambda_1}
\end{equation}
is determined by integral from the  product of the distribution function the Pauli matrices 
in the band representation.
In neglect off-diagonal terms of distribution function we obtain
\begin{equation}
{\bf M}=\int\frac{d^3{\bf k}}{(2\pi)^3}\left [ \frac{\mbox{\boldmath$\gamma$}_{\bf k}}{\gamma}(g_{+}  -g_{-})
\right ].
\end{equation}
Substituting solutions Eq.(\ref{sol})
we obtain
\begin{equation}
{\bf M}=-e\int\frac{d^3{\bf k}}{(2\pi)^3}\frac{\mbox{\boldmath$\gamma$}_{\bf k}}{\gamma}\left [ \tau_+\frac{\partial n_+}{\partial \xi_+}({\bf v}_+{\bf E})-
\tau_-\frac{\partial n_-}{\partial \xi_-}({\bf v}_-{\bf E})
\right ].
\end{equation}
Thus, an application of electric field to a noncentrosymmetric metal causes the appearance the specimen magnetisation.

\subsection{Anomalous Hall effect}

The Hall conductivity is the antisymmetric non-dissipative part of the conductivity tensor $\sigma_{ij}=-\sigma_{ji}$, which determines the relationship between the current and the Hall electric field arising in a magnetic field in a direction perpendicular to both the current and the magnetic field.
\begin{equation}
j_x=\sigma_{xy}E^H_y.
\end{equation}
The anomalous Hall effect arises because, in general, electron states adapt to the presence of an electric field, and the velocity of an electron in a state with momentum ${\bf k}$ acquires an additional term \cite{Chang1995,Xiao2010}
\begin{equation}
{v}_i^n=\frac{1}{\hbar}\frac{\partial\varepsilon^n_{\bf k}}{\partial k_i}+\frac{e}{\hbar}{\Omega}^n_{ij}E_j,
\end{equation}
where ${ \Omega}^n_{ij}$ is the Berry curvature tensor of the $n$-th band with energy $\varepsilon^n_{\bf k}$. Corresponding Hall conductivity is
\begin{equation}
\sigma_{ij}=\frac{e^2}{\hbar}\sum_n\int\frac{d^3{\bf k}}{(2\pi)^3} n(\varepsilon_n){\Omega}^n_{ij}.
\end{equation}
Here $n(\varepsilon_n)=\left \{\exp(\varepsilon_n-\mu)/T+1  \right \}^{-1}$ is the Fermi-Dirac distribution function.

The antisymmetric tensor of Berry curvature for the band $\lambda=+$ is
\begin{eqnarray}
\Omega^+_{ij}=i\left(\frac{\partial \Psi_\alpha^{+\star}}{\partial k_i}\frac{\partial \Psi_\alpha^{+}}{\partial k_j}
-
\frac{\partial \Psi_\alpha^{+\star}}{\partial k_j}\frac{\partial \Psi_\alpha^{+}}{\partial k_i}\right).
\label{om}
\end{eqnarray}
The corresponding Berry curvature for the band $\lambda=-$ is $\Omega^-_{ij}=-\Omega^+_{ij}$, hence, the Hall conductivity is
\begin{equation}
\sigma_{ij}=\frac{e^2}{\hbar}\int\frac{d^3{\bf k}}{(2\pi)^3} \left[n(\varepsilon_+)-n(\varepsilon_-)\right]{\Omega}^+_{ij}.
\label{sigma}
\end{equation}
Let us calculate the Berry curvature for altermagnet with spectrum (\ref{alt1}) in presence of magnetic field along $\hat z$-direction such that 
\begin{eqnarray}
\mbox{\boldmath$\gamma$}_{\bf k}=\gamma_1 \sin(k_zb) \left[\sin(k_ya)\hat x+\sin(k_xa)\hat y\right]\nonumber\\
+(\gamma_2\sin (k_xa)\sin(k_ya)-\mu_BH)\hat z.
\label{alt2}
\end{eqnarray}
Substituting eigen functions (\ref{ps}) in equation (\ref{om}) and performing differentiation we obtain
\begin{eqnarray}
\Omega_{xy}^+=-\frac{\gamma_1^2a^2}{2\gamma^3}\cos(k_xa)\cos(k_ya)\sin^2(k_zb)\nonumber\\
\times[\gamma_2\sin(k_xa)\sin(k_ya)-\mu_bH].
\label{ome}
\end{eqnarray}
Substitution of this expression to Eq.(\ref{sigma}) yields the anomalous Hall conductivity
\begin{eqnarray}
\sigma_{xy}=\frac{e^2\mu_B\gamma_1^2a^2}{2\hbar}H~~~~~~~~~~~\nonumber\\
\times\int\frac{d^3{\bf k}}{(2\pi)^3} \left[n(\varepsilon_+)-n(\varepsilon_-)\right]
\frac{\cos(k_xa)\cos(k_ya)\sin^2(k_zb)}{\gamma^3}.
\label{eq} 
\end{eqnarray}
Similar expression for the Hall conductivity can be found also for noncentrosymmetric metals where  vector $\mbox{\boldmath$\gamma$}_{\bf k}$
is an odd function of the wave vector.

The field independent part of $\Omega_{xy}^+$ vanishes at integration and does not give contribution to the Hall conductivity. This is also true for the Hall conductivity determined by the interband Berry curvature considered in the paper \cite{Mineev2025}.
However, in general one can expect existence of the Hall conductivity even in absence of magnetic field. The possibility of the Hall effect in non-collinear antiferromagnetic materials in the absence of an external magnetic field
was predicted for Mn$_3$Ir about a decade ago \cite{Chen2014,Kubler2014}. Recently, the existence of the same phenomenon in the collinear antiferromagnet RuO$_2$ was pointed out \cite{Smejkal2020}. Numerical estimates of the Hall conductivity in \cite{Chen2014,Kubler2014} as well as in \cite{Smejkal2020} were made using first-principles calculations of the electronic structure. At the moment the corresponding phenomenological derivation is absent. One can only assume that this is achievable making use $4\times 4$ phenomenological spectrum given by Eq.(\ref{four}).

\subsection{Piezomagnetic Hall effect}

Altermagnetics possess piezomafnetism. 
For instance, under stress along diagonal $xy$-direction  an altermagnet with symmetry  (\ref{sym1}) acquires magnetisation along $z$-axis
\begin{equation}
M_z=\lambda_2\sigma_{xy}.
\end{equation}
Hence, an electric current  in basal plane of such altermagnet in presence of $\sigma_{xy}$ stress will induce the appearence 
an electric field in direction perpendicular to the current 
\begin{equation}
{\bf j}_x=\sigma^H_{xy}E_y,
\end{equation}
where the  Hall conductivity 
\begin{eqnarray}
\sigma^H_{xy}=\frac{e^2\mu_B\gamma_1^2a^2}{2\hbar}\lambda_2\sigma_{xy}~~~~~~~~~~~~\nonumber\\
\times\int\frac{d^3{\bf k}}{(2\pi)^3} \left[n(\varepsilon_+)-n(\varepsilon_-)\right]\frac{\cos(k_xa)\cos(k_ya)\sin^2(k_zb)}{\gamma^3}~
\label{eq2} 
\end{eqnarray}
is proportional to the stress.

\section{Superconducting states }

The  BCS  Hamiltonian for singlet pairing in the  spinor representation has the following form 
\begin{eqnarray}
    \hat H=\sum_{{\bf k}\alpha\beta}(\varepsilon_{\bf k}\delta_{\alpha\beta}+\mbox{\boldmath$\gamma$}_{\bf k}\mbox{\boldmath$\sigma$}_{\alpha\beta}-\mu\delta_{\alpha\beta})
   a^\dagger_{{\bf k}\alpha}a_{{\bf k}\beta} ~~~~~\nonumber\\+  
    \frac{1}{2}\sum\limits_{{\bf k}{\bf k}'}\sum_{\alpha\beta\gamma\delta}
  V({\bf k},{\bf k}')(i\sigma_y)_{\alpha\beta}(\sigma_y)^\dagger_{\gamma\delta} 
      a^\dagger_{-{\bf k}\alpha}a^\dagger_{{\bf k}\beta}a_{{\bf k}'\gamma}a_{-{\bf k}'\delta}.~~
\end{eqnarray}
Here
\begin{equation}
V({\bf k},{\bf k}')=-V_0\varphi^\Gamma_i({\bf k})\varphi^{\Gamma\star}_i({\bf k}'),
\end{equation}
is the paring potential decomposed over basis of even $\varphi^\Gamma_i({\bf k})=\varphi^\Gamma_i(-{\bf k})$  functions of given irreducible representation $\Gamma$ of the crystal symmetry group. For example, for altermagnet with symmetry group ${\bf D}_{4h}({\bf D}_{2h})$ consisting of operations enumerated in Eq.(\ref{sym1}) the
function transforming according to unit representation is 
\begin{equation}
\varphi({\bf k})\propto~i(\hat k_x^2-\hat k_y^2).
\end{equation}
Here $\hat k_x,\hat k_y$ are the components of unit vector ${\bf k}/k_F$.

Transforming  to the band representation
\begin{equation}
a_{{\bf k}\alpha}=\Psi^\lambda_\alpha({\bf k})c_{{\bf k}\lambda}
\end{equation}
we obtain
\begin{eqnarray}
\hat H=\sum_{{\bf k}\lambda}(\varepsilon_{\lambda}({\bf k})-\mu)
   c^\dagger_{{\bf k}\lambda}c_{{\bf k}\lambda} ~~~~~~~~~~~\nonumber\\+
   \frac{1}{2}\sum\limits_{{\bf k}{\bf k}'}\sum_
   {\lambda_1\lambda_2\lambda_3\lambda_4}
V_{\lambda_1\lambda_2\lambda_3\lambda_4}({\bf k},{\bf k}') 
      c^\dagger_{-{\bf k}\lambda_1}c^\dagger_{{\bf k}\lambda_2}c_{{\bf k}'\lambda_3}c_{-{\bf k}'\lambda_4}.~~~
\end{eqnarray}
Here,
\begin{equation}
V_{\lambda_1\lambda_2\lambda_3\lambda_4}({\bf k},{\bf k}')=V({\bf k},{\bf k}')t_{\lambda_2}({\bf k})t_{\lambda_4}^\star({\bf k}^\prime)\sigma^x_{\lambda_1\lambda_2}\sigma^x_{\lambda_3\lambda_4} ~~~~
\end{equation}
and $t_\lambda({\bf k})=-\lambda\frac{\gamma_-}{\sqrt{\gamma_+\gamma_-}}$ is the phase factor.

 It is obvious from  this expression that pairing in altermagnets is the pairing of electrons from different bands. This distinguishes them from noncentrosymmetric metals where  
 \begin{equation}
V_{\lambda_1\lambda_2\lambda_3\lambda_4}({\bf k},{\bf k}')=V({\bf k},{\bf k}')t_{\lambda_2}({\bf k})t_{\lambda_4}^\star({\bf k}^\prime)\delta_{\lambda_1\lambda_2}\delta_{\lambda_3\lambda_4} 
\end{equation}
and the pairing  mostly occurs between the electrons from the same band \cite{Samokhin2008}.
 
 The situation in altermagnets reminds pairing in conventional superconductors with singlet pairing in magnetic field which splits the Fermi surfaces for electrons with opposite spins. That leads to paramagnetic suppression of superconductivity. In altermagnets  the same effect takes place in a field absence that leads to effective reduction of temperature of transition to superconducting state or even to complete suppression of superconductivity. Thus, the possibility of existence of superconducting altermagnets raises doubts. Nevertheless, for completeness we present here the  theoretical description of superconductivity in altermagnets.

The Gor'kov equations are
\begin{eqnarray}
 \begin{pmatrix} i\omega\delta_{\lambda_1\lambda_2}-H_{\lambda_1\lambda_2}&-\tilde\Delta_{\lambda_1\lambda_2} \\
 -\tilde\Delta^\dagger_{\lambda_1\lambda_2}& i\omega\delta_{\lambda_1\lambda_2}+H_{\lambda_1\lambda_2}\end{pmatrix}
 \begin{pmatrix}G_{\lambda_2\lambda_3}& -\tilde F_{\lambda_2\lambda_3}\\
-\tilde F^\dagger_{\lambda_2\lambda_3}&-G_{\lambda_2\Lambda_3 }\end{pmatrix}\nonumber\\
=\delta_{\lambda_1\lambda_3} \begin{pmatrix}1&0\\0&1\end{pmatrix},~~~~~
\end{eqnarray} 
where 
\begin{equation}
 i\omega\delta_{\lambda_1\lambda_2}-H_{\lambda_1\lambda_2}=\begin{pmatrix}i\omega-\varepsilon_++\mu&0\\0&i\omega+\varepsilon_--\mu\end{pmatrix}
\end{equation}
and the phase factor is absorbed in the expressions for the order parameter and the Gor'kov function:
\begin{equation}
\Delta_{\lambda_1\lambda_2}({\bf k})=t_{\lambda_2}({\bf k})\tilde\Delta_{\lambda_1\lambda_2}({\bf k}),
\end{equation}
\begin{equation}
\tilde\Delta_{\lambda_1\lambda_2}({\bf k})=(\sigma_x)_{\lambda_1\lambda_2}\Delta({\bf k}),
\end{equation}
\begin{equation}
F_{\lambda_1\lambda_2}({\bf k},\omega_n)=
t_{\lambda_2}({\bf k})\tilde F_{\lambda_1\lambda_2}({\bf k},\omega_n),
\end{equation}
where $\omega_n=\pi T(2n+1)$ is the Matsubara frequency.
The self-consistency equation is
\begin{eqnarray}
\tilde\Delta_{\lambda_1\lambda_2}({\bf k})~~~~~~~~~~~~~~~~~~~~~~~\nonumber\\=-\frac{T}{2}\sum_n\sum_{{\bf k}^\prime}V({\bf k},{\bf k}')(\sigma_x)_{\lambda_2\lambda_1}(\sigma_x)_{\lambda_3\lambda_4}\tilde F_{\lambda_3\lambda_4}({\bf k}^\prime,\omega_n),~~~
\end{eqnarray}
where 
\begin{eqnarray}
\tilde F_{\lambda_1\lambda_2}({\bf k},\omega_n)~~~~~~~~~~~~~~~~~~~~~~~~~~~~~~\nonumber\\= \Delta\begin{pmatrix}0&G_+^n({\bf k},\omega_n)G_-({\bf k},-\omega_n)\\
 G^n_-({\bf k},\omega_n)G_+({\bf k},-\omega_n)&0
 \end{pmatrix}~~~
\end{eqnarray}
is the matrix Gor'kov function and
\begin{equation}
G_\pm({\bf k},\omega_n)=-\frac{i\omega_n+\varepsilon_\pm-\mu}{\omega_n^2+(\varepsilon_\pm-\mu)^2+\Delta^2},
\end{equation}
\begin{equation}
G^n_\pm({\bf k},\omega_n)=\frac{1}{i\omega_n-\varepsilon_\pm+\mu}
\end{equation}
are the band Green functions  in superconducting and normal state correspondingly.
The order parameter in the spin and band representations are related to each other as
\begin{equation}
\Delta_{\alpha\beta}({\bf k})=(i\sigma_y)_{\alpha\beta}\Delta({\bf k}).
\end{equation}

\end{document}